\documentclass{aa}

\usepackage[varg]{txfonts}
\usepackage{epsf}
\usepackage{graphicx}
\usepackage{natbib}
\usepackage{textcomp}
\usepackage{color}
\bibpunct{(}{)}{;}{a}{}{,}

\newcommand{\kep}{{\em Kepler}}

\newcommand{\kepmi}{{\em Kepler Mission}}
\newcommand{\msol}{\ensuremath{\rm{M}_{\odot}}}

\newcommand{\teff}{\ensuremath{T_{\rm{eff}}}}
\newcommand{\logg}{\ensuremath{\log g}}

\newcommand{\lheh}{\ensuremath{\log \left(N_{\mathrm{He}}/N_{\mathrm{H}}\right)}}

\newcommand{\twom}{{\sc 2mass}}

\newcommand{\sigft}{\ensuremath{\sigma_{\mathrm{FT}}}}

\newcommand{\mkep}{{\em Kp}}

\newcommand{\ellone}{\ensuremath{\ell}\,=\,1}
\newcommand{\elltwo}{\ensuremath{\ell}\,=\,2}
\newcommand{\emzero}{\ensuremath{m}\,=\,0}
\newcommand{\emone}{\ensuremath{m}\,=\,\ensuremath{\pm}1}

\newcommand{\DP}{\ensuremath{\Delta P}}
\newcommand{\DPr}{\ensuremath{\Delta\mit\Pi}}
\newcommand{\uHz}{\textmu Hz}

\def\apjl{ApJ}%
\def\aap{A\&A}%

\newcommand{\tlusty}{{\sc tlusty}}
\newcommand{\xtgrid}{{\sc XTgrid}}
\newcommand{\synspec}{{\sc synspec}}
\newcommand{\target}{KIC\,10553698}
\newcommand{\tspan}{855.6\,d}
\newcommand{\npk}{162}
\newcommand{\npid}{156}
\newcommand{\hamfast}{\object{\target A}}
\newcommand{\bell}{\object{\target B}}
\newcommand{\mungo}{\object{KIC\,11558725}}
\newcommand{\bilbo}{\object{KIC\,10670103}}

\newcommand{\pippin}{\object{KIC\,2697388}}
\newcommand{\rosie}{\object{KIC\,5807616}}
\newcommand{\ltarget}{2MASS J19530839+4743002}
\newcommand{\mainmode}{\ensuremath{f_{79-81}}}

\defcitealias{ostensen11b}{{Paper~\sc{i}}}
\defcitealias{baran11b}{{Paper~\sc{ii}}}
\defcitealias{reed11c}{{Paper~\sc{iii}}}

\idline{AA/2014/23611}

\begin{document}

\title{Asteroseismology revealing trapped modes in KIC\,10553698A\thanks{
Based on observations obtained by the \kep\ spacecraft, the Kitt
Peak Mayall Telescope, the Nordic Optical Telescope, and the William
Herschel Telescope.}
}

\author{
   R.~H.~{\O}stensen \inst{1} \and
   J.~H.~Telting     \inst{2} \and
   M.~D.~Reed        \inst{3} \and
   A.~S.~Baran       \inst{4} \and
   P.~Nemeth         \inst{1,5} \and
   F.~Kiaeerad       \inst{2}
}

\institute{
Instituut voor Sterrenkunde, KU\,Leuven, Celestijnenlaan 200D, 3001 Leuven, Belgium\\
   \email{roy@ster.kuleuven.be}
\and Nordic Optical Telescope, Rambla Jos\'e Ana Fern\'andez P\'erez 7, 38711 Bre\~na Baja, Spain
\and Department of Physics, Astronomy, and Materials Science, Missouri State University, Springfield, MO 65804, USA
\and Uniwersytet Pedagogiczny w Krakowie, ul.~Podchor\c{a}\.zych 2, 30-084 Krak\'ow, Poland
\and Dr. Karl Remeis-Observatory \& ECAP, Astronomisches Inst., FAU Erlangen-Nuremberg, 96049 Bamberg, Germany
}

\date{Received 10 February 2014 / Accepted 12 June 2014 }
\abstract{
The subdwarf-B pulsator, \hamfast, is one of 16 such objects observed
with one-minute sampling for most of the duration of the \kepmi.
Like most of these stars, it displays a rich $g$-mode pulsation spectrum with several
clear multiplets that maintain regular frequency splitting. We identify these
pulsation modes as components of rotationally split multiplets in a star
rotating with a period of $\sim$41\,d.
From \npk\ clearly significant periodicities, we are able to identify \npid\ as
likely components of \ellone\ or \elltwo\ multiplets. For the first time we are
able to detect \ellone\ modes that interpose in the asymptotic period sequences
and that provide a clear indication of mode trapping in a stratified
envelope, as predicted by theoretical models.
A clear signal is also present in the \kep\ photometry at 3.387\,d.
Spectroscopic observations reveal
a radial-velocity amplitude of 64.8\,km/s. We find that the
radial-velocity variations and the photometric signal have phase and amplitude
that are perfectly consistent with a Doppler-beaming effect
and conclude that the unseen companion, \bell,\ must be a white dwarf most likely with
a mass close to 0.6\,\msol.
}


\titlerunning{KIC\,10553698: An sdBV+WD binary}
\authorrunning{\O stensen et al.}

\maketitle

\section{Introduction}

Most hot subdwarf-B (sdB) stars belong to the population of
extreme-horizontal-branch (EHB) stars. 
The HB designation implies that they have
ignited helium through the core-helium flash, and therefore
have a core mass close to the helium-flash
mass of $\sim$0.46\,\msol. But in order to reach B-star effective temperatures,
they must have shed almost their entire hydrogen-rich envelope
close to the tip of the red-giant branch.
Several binary scenarios have been identified
that can produce EHB stars, including either common-envelope ejection or
stable Roche-lobe overflow \citep[see][for a detailed review]{heber09}.

Like the main-sequence-B stars, the sdBs pulsate with short ($p$\,modes)
and long periods ($g$\,modes) because of the iron-group elements opacity bump
($\kappa$ mechanism), but at much shorter periods than their main-sequence
counterparts.
A key element in the driving of pulsations in these stars is the competition
between radiative levitation and gravitational settling, which causes a local
overabundance of iron in the driving zone
\citep{charpinet97,fontaine03}.
The hotter short-period pulsators are known as V361-Hya stars after
the prototype \citep{kilkenny97}, and equivalently the cooler long-period pulsators
are known as V1093-Her stars \citep{green03}, and they are collectively referred
to as sdBV stars. Most stars are predominantly
one or the other type, but a few stars in the middle of the temperature range
are hybrids that show both types of pulsations in equal measure
\citep[][and references therein]{ostensen09}.

The \kep\ spacecraft spent four years monitoring a 105\,deg$^2$ field in the Cygnus--Lyrae
region, with the primary goal of detecting transiting planets \citep{borucki11}.
The high-quality lightcurves obtained by the spacecraft reveal a host of
variable stars, providing a treasure trove for asteroseismic studies
\citep{gilliland10a}.
In the first four quarters of the \kepmi, a survey for pulsating stars
was made, and a total of 113 compact-pulsator candidates were checked for
variability by {\O}stensen et al.~(\citealt{ostensen10b}, \citealt{ostensen11b}\,=\,Paper\,{\sc i}).
This very successful survey revealed one clear V361-Hya pulsator \citep{kawaler10a}
and one other transient short-period pulsator, and
a total of thirteen V1093-Her stars \citep[][Paper\,{\sc ii}]{reed10a,kawaler10b,baran11b},
including an sdB+dM eclipsing binary in which the hot primary
shows an exceptionally rich pulsation spectrum \citep{2m1938}. Another three
V1093-Her pulsators have been identified in the open cluster NGC\,6791
\citep{pablo11,reed12b}, bringing the total number of sdBV stars in the \kep\ field
to eighteen. A closely related object was discovered in \kep-archive data by \citet{ostensen12b};
a BHB pulsator showing pulsation properties similar to the V1093-Her pulsators -- the
first such object discovered.

\begin{figure*}[t!]
\centering
\includegraphics[width=14cm]{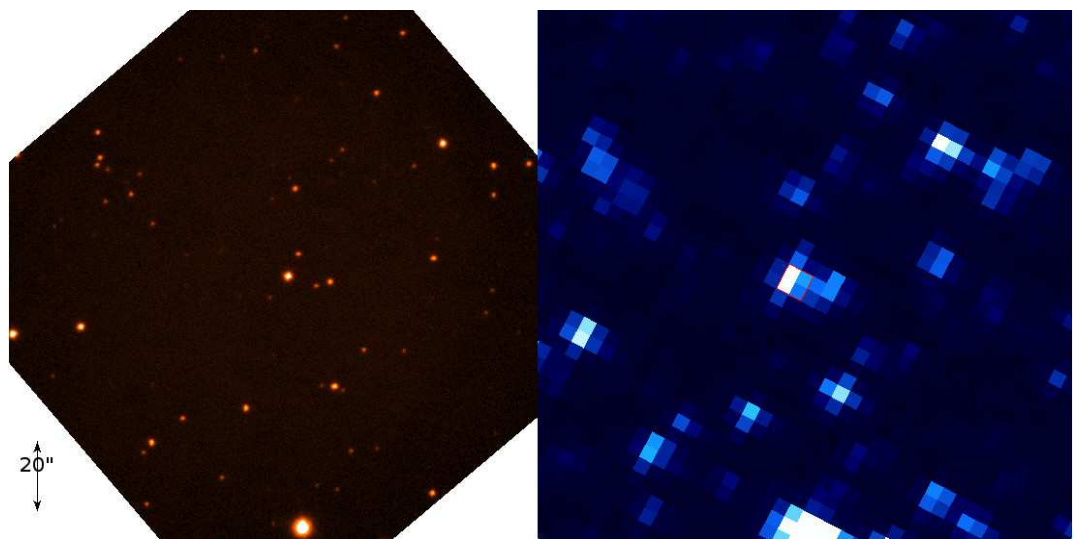}
\caption{
Field images for \target. The NOT/ALFOSC acquisition image (left) covers
2.5$\times$2.5 arc minutes,
and the corresponding section from a \kep\ full-frame image is also shown (right).
The images are aligned so that north is up and east to the left.
The target is the bright star in the centre of the field, and the four central
pixels that are used for photometry for a typical quarter are outlined in the
\kep\ image.
}
\label{fig:field}
\end{figure*}

Since the series of early papers based on one-month datasets obtained during
the survey phase of the \kepmi, only
seven of the $g$-mode pulsators have been subjected
to detailed analysis based on many months of near-continuous data that the \kep\ spacecraft 
gathered during the long-term monitoring phase.
First, \citet{charpinet11b} analysed one year of data on \rosie,
revealing long-period periodicities that may be signatures of planets in very close
orbit around the otherwise single sdB star.
\citet{baran12b} analysed nine months of data on \pippin, and
\citet{telting12a} provided a detailed study of the 10-day sdBV+WD binary
\mungo, based on 15 months of data. The study of \citet{baran12c}
analysed 27 months of data on KIC\,2991403, KIC\,2438324, and KIC\,11179657.
Most recently, \citet{reed14} have analysed the full 2.75 year dataset of \bilbo,
providing mode identifications for 178 of 278 detected frequencies.
While the papers by \citet{VanGrootel10} and \citet{charpinet11a}
matched asteroseismic models to observed frequencies based on survey data,
no paper has yet attempted such forward modelling on a full two- to three-year
\kep\ dataset.

The target presented in this work, \target, was included in the original survey
and identified as a $g$-mode pulsator in \citetalias{ostensen11b}.
The one-month discovery run was examined in \citetalias{baran11b}
where 43 pulsation frequencies were identified. Thirty-seven of those frequencies were
clearly in the $g$-mode region between 104 and 493\,\uHz, four were found in
the intermediate region between 750 and 809\,\uHz, and two were in the
high-frequency $p$-mode region at 3074 and 4070\,\uHz.
\citet[][Paper\,{\sc iii}]{reed11c} analysed the period spacing in this star along with
thirteen other $g$-mode pulsators and made a first estimate of the mean period spacings
for the non-radial $g$ modes of degree \ellone\ and \elltwo.

Since those early \kep\ papers,
\target\ was monitored by \kep\ throughout quarters 8 to 17, but
skipping Q11 and Q15 when the object fell on CCD Module \#3, which failed in
January 2010.
It is one of the brighter V1093-Hya stars in the sample, but it suffers
some contamination in the \kep\ photometry due to crowding from nearby stars,
as illustrated in Fig.\,\ref{fig:field}.
Here we analyse all the available quarters of data from Q8 through to the end 
of the mission in Q17, and we
provide a detailed frequency analysis with identification of the modes detected
in the frequency spectrum.
We also present new spectroscopic observations that demonstrate that \target\
is a single-lined spectroscopic binary with a 3.6\,d orbital period. We
refer to the pulsator as \hamfast and the invisible companion as \target B.

\section{Spectroscopic observations}

We observed \target\ as part of our observing campaign dedicated to
investigating the binary status of the hot subdwarfs in the \kep\ field
\citep{telting12b,telting13}. From 2010 to 2012 we obtained 42
radial-velocity (RV) measurements of \target, as listed in
Table~\ref{tbl:rvs}. The first seven observations were made on each of
six consecutive nights between August 25 and 30, 2010, using the ISIS
long-slit spectrograph on the William Herschel Telescope (WHT), equipped
with the R600B grating and using a 1.0" slit. It was already clear from 
this initial set of high signal-to-noise (S/N) observations that the target changes its
RV with a peak-to-peak amplitude of $\sim$130\,km/s with a period of
around three days.

Additional spectroscopy were collected with the ALFOSC spectrograph at
the Nordic Optical Telescope (NOT)
between May 2011 and October 2012. In total 24 spectra were collected
over 18 nights using an 0.5" slit with grism\,\#16. The final four of these
were obtained using the spectrograph in vertical slit mode for fast
readout. We found that these observations appear to be systematically
shifted by $\sim$30 km/s with respect to the other observations,
indicating a problem with the wavelength calibration in this setup. These
points were discarded rather than attempting to correct for the unexpected
offset. A final set of ten spectra were obtained between September 29 and
October 1, 2012, with the Kitt Peak 4-m Mayall telescope (KPNO) with RC-Spec/F3KB,
the kpc-22b grating and a 2" slit. The WHT dataset is clearly the best
with a S/N level close to 100\ in each spectrum.
The NOT data have variable quality between S/N\,$\sim$\,20 and 80
depending on observing conditions, and the KPNO data have S/N\,$\sim$\,50.

All spectra were processed and extracted using standard
{\scriptsize IRAF}\footnote{
{\sc iraf} is distributed by the National Optical Astronomy Observatory;
see http://iraf.noao.edu/.} tasks. Radial velocities (RVs)
were computed with {\tt fxcor}, by cross-correlating with
a synthetic template derived from a fit to a mean spectrum of the target, and
using the $H\gamma$, $H\delta$, $H\zeta$, and $H\eta$ lines.
For the ALFOSC data, the final RVs were adjusted for the position of
the target on the slit, judged from slit images taken just before and
after the spectra. 
Table~\ref{tbl:rvs} lists the observations with their mid-exposure dates,
and RV measurements with the {\tt fxcor} error ({\sc verr}), as well as
the observatory tabulated in the last column.

For the final determination of the RV amplitude we used the orbital
period and phase as determined from the \kep\ photometry
(see Section\,\ref{sect:orbit}, below), since this can be determined 
much more accurately than from the sparse spectroscopic observations.
Fitting the amplitude and systemic velocity, we find
\begin{eqnarray*}
K_1 & = & 64.8 \pm 2.2\,\mathrm{km/s} \\
\gamma & = & 52.1 \pm 1.5\,\mathrm{km/s}.
\end{eqnarray*}
The phase-folded RV measurements are plotted together with the photometric
orbital signal in Fig.\,\ref{fig:foldplot}.
The mass function is then
\begin{equation}\label{eq:fm}
f(m) = \frac{(M_2 \sin i)^3}{(M_1 + M_2)^2} = 1.036\cdot10^{-7} K_1^3 P,\msol = 0.095\,\msol,
\end{equation}
which for a canonical 0.46\,\msol\ primary, provides a minimum mass
for the secondary of 0.42\,\msol.
Thus, unless the orbital inclination is
less than 29$^\circ$, \bell\ must be a white dwarf (WD).
For a canonical 0.6\,\msol\ WD, the inclination angle is $\sim52^\circ$.

\begin{table}[b]\small\rm
\caption[]{Physical parameters derived from the detrended mean spectra.}
\label{tbl:physpar}
\centering
\begin{tabular}{llll} \hline\hline \noalign{\smallskip}
Spectrum   & \teff & \logg & \lheh \\
   & [K]  & [dex] & [dex] \\
\noalign{\smallskip} \hline \noalign{\smallskip}
WHT        & 27413\,$\pm$\,\ 67 & 5.461\,$\pm$\,0.011 & --2.838\,$\pm$\,0.024 \\
NOT1       & 27007\,$\pm$\,136  & 5.404\,$\pm$\,0.021 & --2.809\,$\pm$\,0.017 \\
KPNO       & 27712\,$\pm$\,\ 90 & 5.425\,$\pm$\,0.016 & --2.792\,$\pm$\,0.024 \\
\noalign{\smallskip} \hline \noalign{\smallskip}
Adopted    & 27423\,$\pm$\,293  & 5.436\,$\pm$\,0.024 & --2.813\,$\pm$\,0.019 \\
\noalign{\smallskip} \hline
\end{tabular}
\end{table}

\begin{figure}
\includegraphics[width=\hsize]{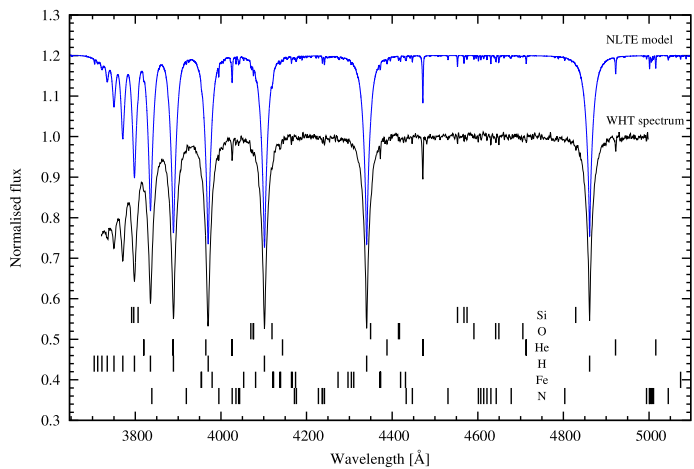}
\caption{
Mean WHT spectrum of KIC\,10553698 after correcting for the orbital velocity.
The S/N in this mean spectrum peaks at $\sim$200, and many weak metal lines
can be distinguished in addition to the strong Balmer lines and \ion{He}{i}
lines at 4472 and 4026\,\AA.
The continuum of the normalised spectrum was sampled individually in 100\,\AA\
sections.
Shifted up by 0.2 is the model fit
computed with \tlusty/\xtgrid. Line identifications are given for lines
stronger than 30\,m\AA\ in the model. The final parameters for this model fit are given
in Table\,\ref{tbl:abund}. 
}
\label{fig:wht_sp} 
\end{figure}

\subsection{Atmospheric properties of the sdB}

We determined \teff\ and \logg\ from each of three mean spectra,
co-added after correcting for the radial-velocity variation of the
orbit. We redetermined the physical parameters of \hamfast, using the
H/He LTE grid of \citet{heber00} for consistency with \citetalias{ostensen11b}.
We used all the Balmer lines from $H\beta$ to $H\kappa$ (excluding only
the $H\epsilon$ line due to contamination with the \ion{Ca}{ii}-H line) and
the five strongest \ion{He}{i} lines for the fit. The results are listed
in Table~\ref{tbl:physpar}, with the error-weighted mean in the bottom row.
The errors listed on the measurements are the formal errors of the fit, which reflect
the S/N of each mean, while
the errors on the adopted values are the rms of the three measurements, which
reflect the systematics of using different spectrographs more than the quality
of the observations. 
These values and errors are relative to the LTE model grid and do not
reflect any systematic effects caused by the assumptions underlying those models.

\begin{table}[b!]
\caption[]{Fitted lines with equivalent widths larger then 50\,m\AA.}
\label{tbl:lines}
\centering
\begin{tabular}{lcr|lcr} \hline\hline \noalign{\smallskip}
Ion   & Wavelength & \multicolumn{1}{c}{$W_\lambda$} & Ion   & Wavelength  & \multicolumn{1}{c}{$W_\lambda$}  \\
      & [\AA]      & [m\AA] & & [\AA]      & [m\AA] \\
\noalign{\smallskip} \hline \noalign{\smallskip}
He {\sc i}  & 3819.60  & 67.7  & N {\sc ii}  & 4447.03  & 68.5  \\
He {\sc i}  & 3888.60  & 57.6  & N {\sc ii}  & 4530.41  & 80.5  \\
He {\sc i}  & 3888.65  & 228.1 & N {\sc ii}  & 4607.15  & 53.6  \\
He {\sc i}  & 3964.73  & 72.2  & N {\sc ii}  & 4621.39  & 65.7  \\
He {\sc i}  & 4026.19  & 232.0 & N {\sc ii}  & 4630.54  & 95.6  \\
He {\sc i}  & 4387.93  & 98.8  & N {\sc ii}  & 5001.13  & 63.3  \\
He {\sc i}  & 4471.47  & 295.2 & N {\sc ii}  & 5001.47  & 83.2  \\
He {\sc i}  & 4471.49  & 258.9 & N {\sc ii}  & 5005.15  & 96.5  \\
He {\sc i}  & 4471.68  & 79.0  & N {\sc ii}  & 5045.10  & 80.7  \\
He {\sc i}  & 4713.14  & 50.0  & O {\sc ii}   & 4349.43 & 63.8  \\
He {\sc i}  & 4921.93  & 154.4 & O {\sc ii}   & 4641.81 & 55.9  \\
He {\sc i}  & 5015.68  & 96.2  & O {\sc ii}   & 4649.14 & 55.3  \\
N {\sc ii}  & 3994.99  & 95.2  & Si {\sc iii} & 3806.53 & 54.6  \\
N {\sc ii}  & 4041.31  & 68.6  & Si {\sc iii} & 4552.62 & 73.7  \\
N {\sc ii}  & 4043.53  & 59.4  & Si {\sc iii} & 4567.84 & 65.3  \\
N {\sc ii}  & 4237.05  & 71.1  & Fe {\sc iii} & 4137.76 & 50.6  \\
N {\sc ii}  & 4241.79  & 53.0  & Fe {\sc iii} & 4164.73 & 54.1  \\
N {\sc ii}  & 4432.74  & 62.6  & Fe {\sc iii} & 4419.60 & 51.6  \\
\noalign{\smallskip} \hline
\end{tabular}\end{table}   

\begin{table}[b]
\caption[]{NLTE atmospheric parameters for the fit shown in Fig.\,\ref{fig:wht_sp}, with
respect to the solar abundances from \citet{grevesse98} provided for comparison.}
\label{tbl:abund}
\centering
\begin{tabular}{llrrrl} \hline\hline \noalign{\smallskip}  
Parameter                      & Value   & $+1 \sigma$ & $-1 \sigma$ & $\times$\,Solar & Unit\\
\noalign{\smallskip} \hline \noalign{\smallskip}
\teff\                           & 27750 & 130 & 70 & & K\\
\logg\                           &  5.452 & 0.020 & 0.008 & & dex \\
$\log n(\mathrm{He})/n(\mathrm{H})$\ &--2.74 & 0.03 & 0.11 & 0.018 & dex \\
$\log n(\mathrm{C})/n(\mathrm{H}) $\ &--6.1> &      &      & 0.001 & dex \\
$\log n(\mathrm{N})/n(\mathrm{H}) $\ &--4.45 & 0.13 & 0.23 & 0.427 & dex \\
$\log n(\mathrm{O})/n(\mathrm{H}) $\ &--4.63 & 0.31 & 0.18 & 0.035 & dex \\
$\log n(\mathrm{Si})/n(\mathrm{H})$\ &--5.65 & 0.24 & 0.46 & 0.063 & dex \\
$\log n(\mathrm{Fe})/n(\mathrm{H})$\ &--4.30 & 0.36 & 0.05 & 1.580 & dex \\
\noalign{\smallskip} \hline
\end{tabular}
\end{table}

We also fitted the mean WHT spectrum with the NLTE model atmosphere code \tlusty\
\citep{tlusty}.
Spectral synthesis was done with \synspec\,{\small 49}.
Our models included H, He, C, N, O, Si and Fe opacities consistently in the
calculations for atmospheric structure and synthetic spectra.
The fit to the observation was done by the \xtgrid\ fitting program
\citep{nemeth12}. 
This procedure is a standard $\chi^2$-minimisation technique, which starts
with a detailed model, and by successive approximations
along the steepest gradient of the $\chi^2$, it converges on a solution.
Instead of individual lines, the procedure fits the entire spectrum so  
as to account for line blanketing.  
However, the fit is still driven by the dominant Balmer lines with
contributions from the
strongest metal lines (listed in Table~\ref{tbl:lines}).
The best fit was found with \teff\,=\,27750\,K
and \logg\,=\,5.45\,dex, using the Stark broadening tables of
\citet{tremblay09}.  
Errors and abundances for those elements
that were found to be significant are listed in Table\,\ref{tbl:abund}.
When using the VCS Stark broadening tables for hydrogen \citep{lemke97}, we
found a systematically lower
surface temperature and gravity, by 800\,K and 0.06\,dex.
Parameter errors were determined by changing the model in one dimension\
until the critical $\chi^2$-value associated with the probability level at the
given number of free parameters was reached.
The resulting fit is shown together with the mean spectrum
in Fig.\,\ref{fig:wht_sp}.
We used a resolution of $\Delta\lambda=1.7$ \AA\ and assumed a non-rotating
sdB star.

Our NLTE model provides consistent parameters with the LTE analysis,
indicating that NLTE effects are negligible in the atmosphere of \hamfast.
The abundances show that iron is supersolar, nitrogen is about half solar,
whereas the other elements are significantly depleted with respect to their solar abundances.
This pattern fits in the typical abundance pattern of sdB stars
\citep[see e.g.][]{geier13a}.

\begin{figure*}[t!]
\centering
\includegraphics[width=\hsize]{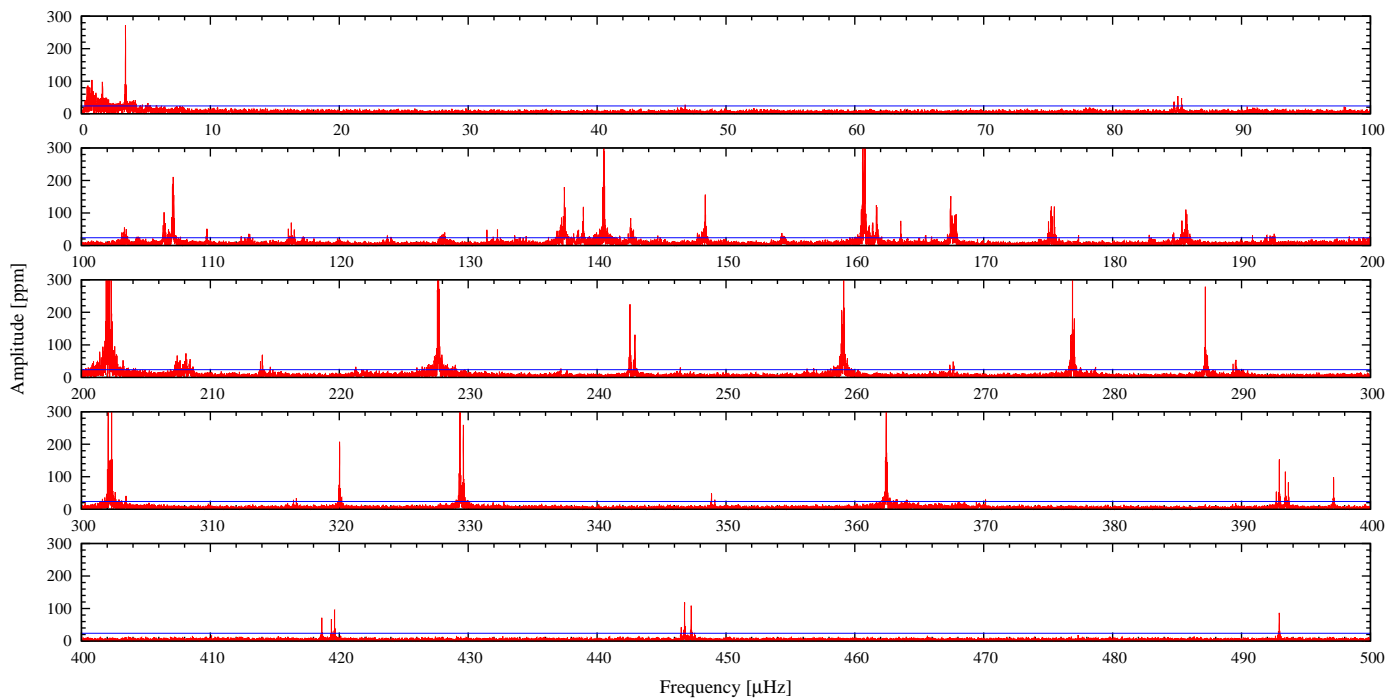}
\caption{
FT of the full \kep\ dataset of \target.
The ordinate axis has been truncated at 300\,ppm to show sufficient
details, even if there are some peaks in the FT that exceed this value.
The 5-$\sigma$ level is indicated by a continuous line.
}
\label{fig:kepft}
\end{figure*}

\section{Photometry and frequency analysis}

\target\ has a magnitude in the \kep\ photometric passband of \mkep\,=\,15.134 and
colours $g-r$\,=\,--0.395 and $g-i$\,=\,--0.694\footnote{The \kep\ Input Catalog
does not provide errors on the magnitudes}. It also
appears in the \twom\ catalogue as \ltarget\ with $J$\,=\,15.45(5), $H$\,=\,15.54(9)
but is below the detection limit in $K_\mathrm{s}$.
Its close proximity to several fainter stars makes the \kep\ photometry
suffer from contamination that varies slightly from quarter to quarter, depending
on the positioning of the instrument's 4"-sized pixels.
Figure\,\ref{fig:field} shows a 150"-sized section of an ALFOSC target acquisition frame
with the corresponding section of a \kep\ full-frame image. 
For the frequency analysis, we used the optimally extracted
lightcurves provided by the MAST\footnote{The Mikulski Archive for
Space Telescopes is hosted by the Space Telescope Science Institute (STScI) at
http://archive.stsci.edu/.}.
These were detrended using low-order polynomials for each
continuous lightcurve segment, removing only trends on month-long timescales. 
We experimented with using the pixel data in order to retain more flux from the target,
but no significant improvement was achieved, so all the data presented here are based
on the standard extraction provided by the archive pipeline.

\begin{figure}[t!]
\includegraphics[width=\hsize]{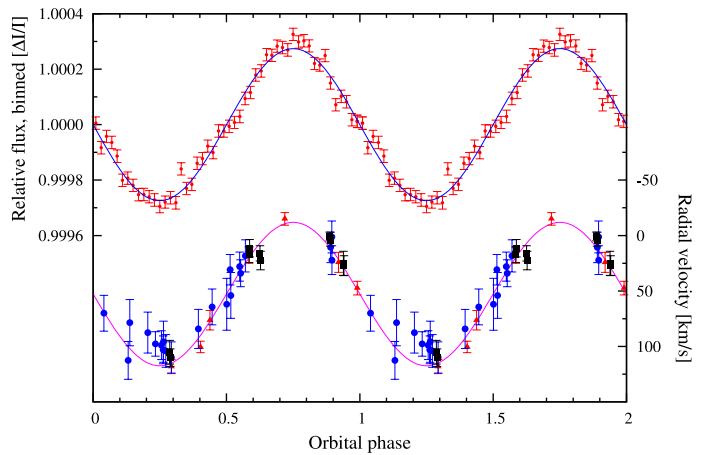}
\caption{
Top: the \tspan\ \kep\ lightcurve folded on the orbital period and binned into 50 bins.
Bottom: radial velocity measurements folded on the same ephemeris as the lightcurve.
(Red triangles: WHT; blue circles: NOT; black squares: KPNO)
}
\label{fig:foldplot}
\end{figure}

\begin{figure*}[t!]
\includegraphics[width=\hsize]{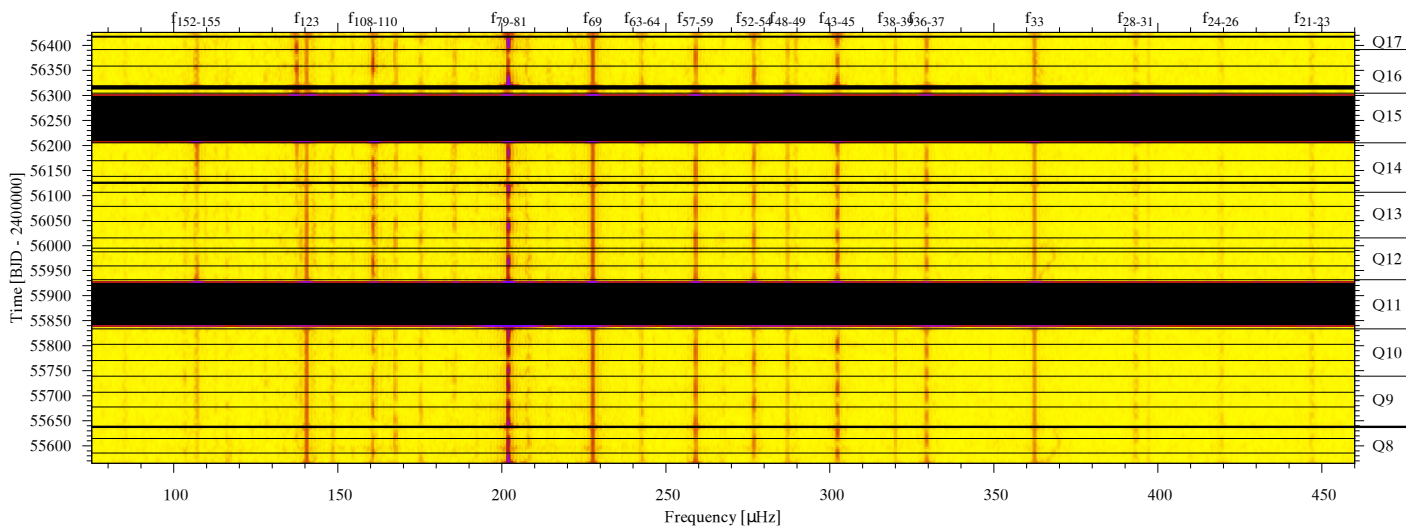}
\caption{
Sliding FT for the region with the most significant pulsations in \target. 
}
\label{fig:running}
\end{figure*}

\subsection{The orbital signal}\label{sect:orbit}

In the Fourier transform (FT) shown in Fig.\,\ref{fig:kepft}, the first significant
peak is found at 3.41678\,\uHz, which corresponds to a period of 3.38743\,d.
Since the span of the \kep\ dataset is \tspan, the frequency resolution
is 0.014\,\uHz, which coincidentally corresponds to a precision in period
of 0.014\,d. Thus, 
assuming a circular orbit, we find an ephemeris for the system
\begin{eqnarray*}
P_\mathrm{orb} & = & 3.387 \pm 0.014\ \mathrm{d},\\
T_0 & = & 55436.468 \pm 0.014\ \mathrm{d},
\end{eqnarray*}
where $T_0$ is the time corresponding to zero phase
(where the subdwarf is at the closest point to the observer)
for the first epoch of observations.
Since the orbital signal is too weak for single minima to be detected in the
point-to-point scatter, the uncertainty
on the ephemeris phase is the same as that of the Fourier analysis. 
Figure\,\ref{fig:foldplot} shows the lightcurve folded on this ephemeris
and binned into 50 points. The error bars reflect the rms noise in these
bins, which may be boosted by the pulsation signal.

The photometric orbital signal seen in the \kep\ lightcurve is caused
by the Doppler-beaming effect, as described in detail for the 0.4-d
eclipsing sdB+WD binary \object{KPD\,1946+4330} by \citet{bloemen11} and the
10-d sdBV+WD binary \object{KIC\,11558725} by \citet{telting12a}. 
\target\ is similar to the latter in that it only displays the beaming
signal and no ellipsoidal deformation, which would be present at
$P_\mathrm{orb}$/2, as is clearly seen in \object{KPD\,1946+4330} and also in
\object{KIC\,6614501} \citep{silvotti12}.
For plain Doppler beaming, the observed flux from the target is related
to the orbital velocity along the line of sight, $v_r$, as
\begin{equation}
F = F_0 \left ( 1 - B \frac{v_r}{c} \right )
\end{equation}
where $F_0$ is the intrinsic flux of the star in the observed passband,
$B$ the beaming factor, and $v_r$/$c$ the fraction of the orbital
velocity to the speed of light. The beaming factor has several
terms, some geometrical, and a part that depends on the spectrum of
the radiating star. Since \hamfast\ has almost exactly the same physical
parameters as \mungo, we simply adopt the beaming factor computed by
\citet{telting12a}, $B$\,=\,1.403(5). 
Using the measured RV amplitude we can then predict a beaming amplitude of
$B K_1/c$\,=\,303\,$\pm$\,10\,ppm.

The observed beaming amplitude is 274\,$\pm$\,6\,ppm in the optimally extracted
lightcurve, after applying minimal polynomial fits to remove long-term trends.
Correcting for the crowding fraction indicated in the \kep\ dataset,
which states that the ratio of target flux to total flux is $\sim$0.91 
(see Table~\ref{tbl:pixpars}),
the contamination-corrected beaming amplitude is 299 ppm, which is perfectly
consistent with the predicted value.

\begin{table}[bt]
\caption[]{\kep\ pixel-data parameters.}
\label{tbl:pixpars}
\centering
\begin{tabular}{lccc} \hline\hline \noalign{\smallskip}
Quarter    & $N_\mathrm{pix}$ & CROWDSAP \tablefootmark{a} & FLFRCSAP \tablefootmark{b} \\
\noalign{\smallskip} \hline \noalign{\smallskip}
Q8, Q12, Q16 & 4 & 0.9079 & 0.7305 \\
Q9, Q13      & 5 & 0.9176 & 0.8128 \\
Q10, Q14     & 6 & 0.9026 & 0.8212 \\
Q17          & 5 & 0.9070 & 0.8087 \\
\noalign{\smallskip} \hline
\end{tabular}
\tablefoot{
\kep\ FITS data file header keywords, indicating:\\
\tablefoottext{a}{Ratio of target flux to total flux in optimal aperture}\\
\tablefoottext{b}{Fraction of target flux within the optimal aperture}\\
No errors are provided on these values.
}
\end{table}

\subsection{The sliding FT}

In Fig.\,\ref{fig:running} we show a sliding FT (sFT) of the same dataset as was used to
generate Fig.\,\ref{fig:kepft}. The data were chopped into segments of 12-d
length and stepped with 4-d intervals, and the resulting FTs are stacked
with time running in the y-direction to visualise the time variability
of the modes.
The black bands indicate the data gaps,
most significantly the Q11 and Q15 gaps, when \target\ fell on the defunct
Module \#3. The thin black lines indicate the regular monthly data-downlink
gaps of typically one-day duration, with slightly thicker lines indicating other
events that caused interruptions to the observations for various reasons.
A spacecraft artefact is seen close to $f_{33}$ at $\sim$370\,\uHz\ in Q8 and recurring
every year, as described by \citet{baran13b}.

It is easy to see from the sFT that some frequencies are single and very stable
(e.g.~$f_{33}$, $f_{69}$, and $f_{123}$), and some produce stable
beat patterns caused by doublets and triplets that are unresolved in the
12\,d chunks (e.g.~\mainmode, $f_{43-45}$, $f_{36-37}$, and $f_{63-64}$).
At higher frequencies, $f_{21-23}$, $f_{24-26}$ and $f_{28-31}$ form broader,
more complex patterns that also appear to be completely stable throughout the
duration of the observations.
In the low-frequency range, the modes seem less stable, sometimes appearing
or disappearing completely. For instance, the mode labelled $f_{108-110}$
appears to have a similar beat period as that of the strongest mode, \mainmode,
but with an amplitude that increases throughout the run duration.

\begin{figure*}[t!]
\centering
\includegraphics[width=\hsize]{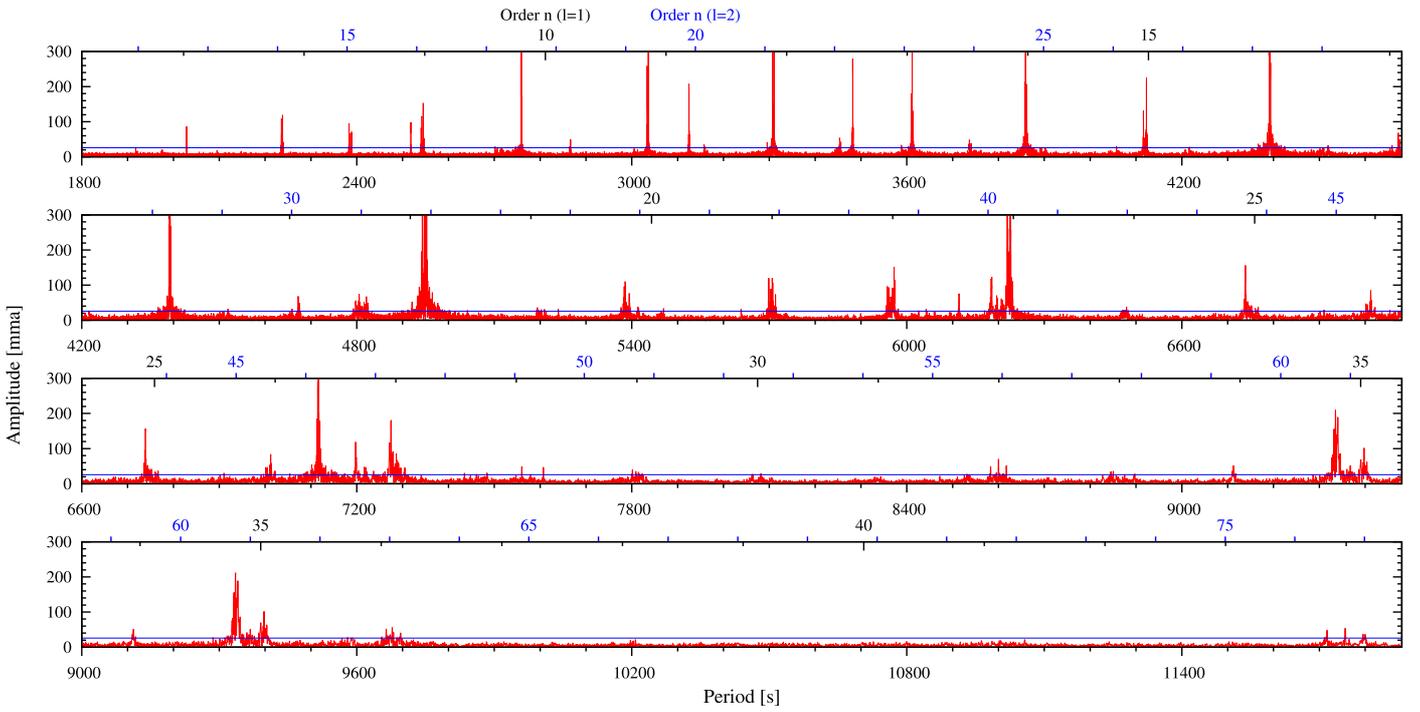}
\caption{ \label{fig:kepper}
Periodogram for \target. This is the same FT as in Fig.\,\ref{fig:kepft},
but with period on the abscissa. The first panel starts where the last panel in 
Fig.\,\ref{fig:kepft} ends, and frequencies shorter than the triplet at
86\,\uHz\ have been truncated. Radial order according to the asymptotic
relation (Eq.~\ref{eq:asym}) is indicated on the top axis, with inward
tick marks counting the \ellone\ spacing and outward tick marks counting
the \elltwo\ spacing.
}
\end{figure*}

\begin{figure}[t!]
\includegraphics[width=8.5cm]{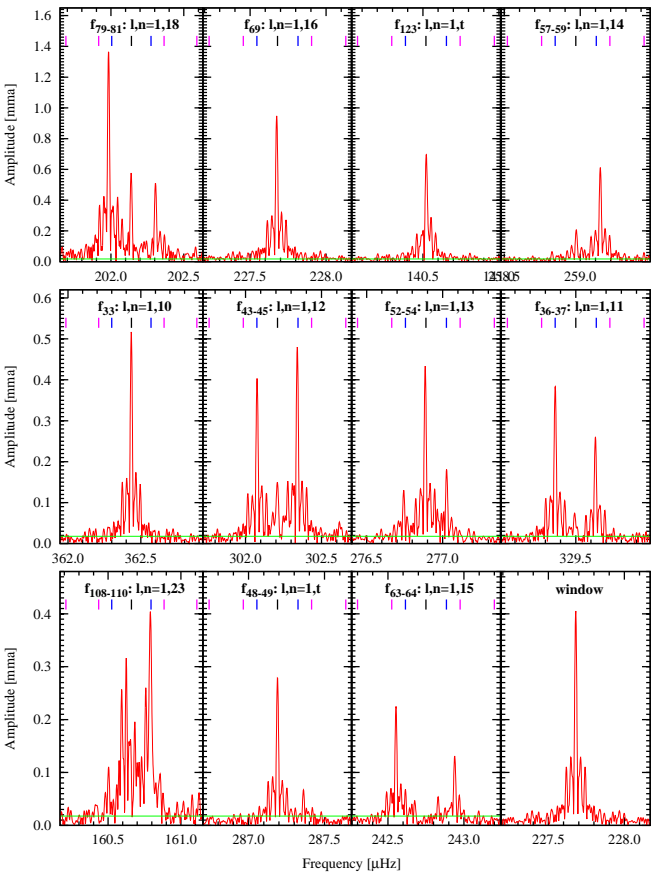}
\caption{ \label{fig:mainpk}
Eleven highest amplitude frequencies in the \target\ Fourier spectrum.
The bars indicate positions for the central \emzero\ component (black),
\ellone, \emone\ (blue), and \elltwo, \emone,2 (magenta).  }
\end{figure}

\subsection{The Fourier spectrum}

A careful analysis of the peaks in the FT of \target\ reveals \npk\ significant
features. What can be considered significant is always a matter of interpretation
in such analyses, especially when amplitude variability is present. 
Here we have only retained frequencies that appear well separated
rather than try to include every peak that appears in a cluster,
since many of them are likely to be caused by splittings produced by
amplitude variability. We analysed both the FT of the full SC dataset, including all seven
available quarters, as plotted in Fig.\,\ref{fig:running}, and the FT of the two
long runs from Q8-10 and Q12-14 separately. We set the detection limit to five times
the mean level in the FT, \sigft, which translates to 25.7\,ppm for the 8-Q dataset and $\sim$40\,ppm
for the 3-Q datasets. To be retained, we required every frequency to be above
the respective 5\sigft\ limit in at least one of these three sets.

The full list of frequencies is provided in the Appendix, Table~\ref{tbl:freqs}.
The table lists the frequency, period, and amplitude of each detected mode, together
with a tentative mode ID provided as a non-radial degree number $\ell$ and a radial
order $n$, where one could be estimated
based on the analysis of the \'echelle diagram discussed in Section\,\ref{sect:modeID},
below.  Also given is a `State' description indicating the
stability of the mode. This is given as `stable', `rising', or `dropping' if the
mode is present in both 3-Q datasets within $\pm$0.05\,\uHz\ of the
frequency detected in the 8-Q dataset, and `stable' to indicate that the amplitude
does not change by more than 20\%\ between the two 3-Q sets compared with the
amplitude of the 8-Q set.
Modes that are not detected in one of the 3-Q datasets are classified as
`appearing' or `disappearing', and modes that are only significant in the 8-Q dataset
are labelled `noisy'. A few modes labelled `messy' are significant only in one of the
3-Q datasets, but because of cancellation effects, they make only a broad, non-significant peak
in the 8-Q set. 

\subsection{Rotationally split triplets}

In Fig.\,\ref{fig:mainpk} we show the eleven strongest peaks
together with the window function. It is quite clear from
this figure already that many peaks appear in groups with a common spacing, as would be
expected for rotationally split multiplets. The picture is not as clear as one may have
wished. The strongest triplet, \mainmode, shows slightly uneven splittings of
$\delta\nu$\,=\,0.168 and 0.155\,\uHz, respectively. The triplet $f_{57-59}$ is perfectly even with
a splitting of 0.159\,\uHz. Both are somewhat lopsided in amplitude, leaning either to
the $m$\,=\,+1 or the $m$\,=\,--1 side. The three symmetric triplets $f_{43-45}$, $f_{52-54}$,
and $f_{36-37}$ have splittings of 0.132, 0.141, and 0.132\,\uHz, respectively.

The difference in
rotational period that can be inferred from these splittings is quite significant 
(when using a simple model with a Ledoux constant,
$C_{n\ell}$,
of 0.5, as expected for high-order g-modes
of degree \ellone), ranging from 34.5 to 44.2\,d.
To indicate the rotational splitting in Figures 7, 9, and B.1, 
we have used an average of the splittings measured in
the three consecutive triplets identified as $n$\,=\,11, 12, and 13,
$\delta\nu$\,=\,0.135\,\uHz, implying a
rotational period of 42.9\,d.

\subsection{Quintuplets and the pulsation axis}

Another spectacular sequence of multiplets can be found at higher frequencies.
The four peaks labelled $f_{28-31}$ at 393\,\uHz\ form a perfectly even quintuplet with the middle
peak missing and a splitting of 0.235\,\uHz. The sequence $f_{24-26}$ at 419\,\uHz\ appears as
three components of a quintuplet with splitting of 0.247\,\uHz, and $f_{21-23}$ likewise matches three
components of a quintuplet with splitting of 0.258\,\uHz. In all three cases there are indications of the
missing components close to the 5\sigft\ limit. The additional fact that this sequence of multiplets
appears with a spacing of $\sim$150\,s\,=\,260\,s/$\sqrt{3}$ (see below)
makes it very clear that this is
a sequence of consecutive \elltwo\ modes.

The rotational splitting for high-order $g$-modes in a uniformly rotating star is given by
\citet{asteroseismology}
\begin{equation}
\delta\nu = m \Omega\,(1 - C_{n\ell}) \simeq \frac{m}{P_\mathrm{rot}}\left( 1 - \frac{1}{\ell(\ell+1)} \right),
\end{equation}
so that the observed splittings of the three consecutive \elltwo\ multiplets
translate to a rotation period of 41.0, 39.0, and 37.4\,d, close to the middle of the range seen
for the \ellone\ modes.

It is interesting that the middle \emzero\ component is suppressed in this sequence of multiplets. 
This is very different from the quintuplet structure seen in \bilbo~\citep{reed14} where the middle
component is the strongest one.
Geometric cancellation of $\ell,m$\,=\,2,0 occurs only when viewing a pulsator within a few degrees
of $i$\,=\,55\degr. At this angle there is no significant suppression of any particular \ellone\
component, consistent with what we have seen.

From the spectroscopic observations, we found that the mass function implies
that the companion is consistent with a white dwarf for all orbital
inclinations higher than 29\degr.
If the pulsation axis is aligned with the orbital axis, the mass
of the white dwarf must be close to 0.6\,\msol, which is the typical value for
white dwarfs that are remnants of intermediate-mass stars after normal 
uninterrupted evolution.

\begin{figure}[t]
\includegraphics[width=\hsize]{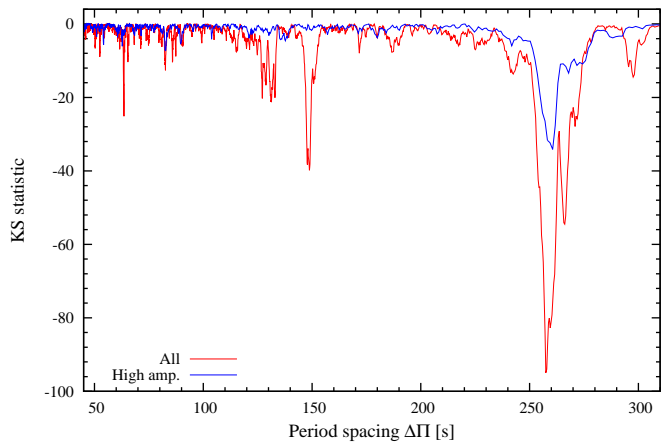}
\caption{\label{fig:kstest}
KS test statistic for the full frequency list (red) and high-amplitude
($A$\,$>$\,1000\,ppm) modes (red) respectively.
}
\end{figure}

\begin{figure*}[t!]
\includegraphics[width=\hsize]{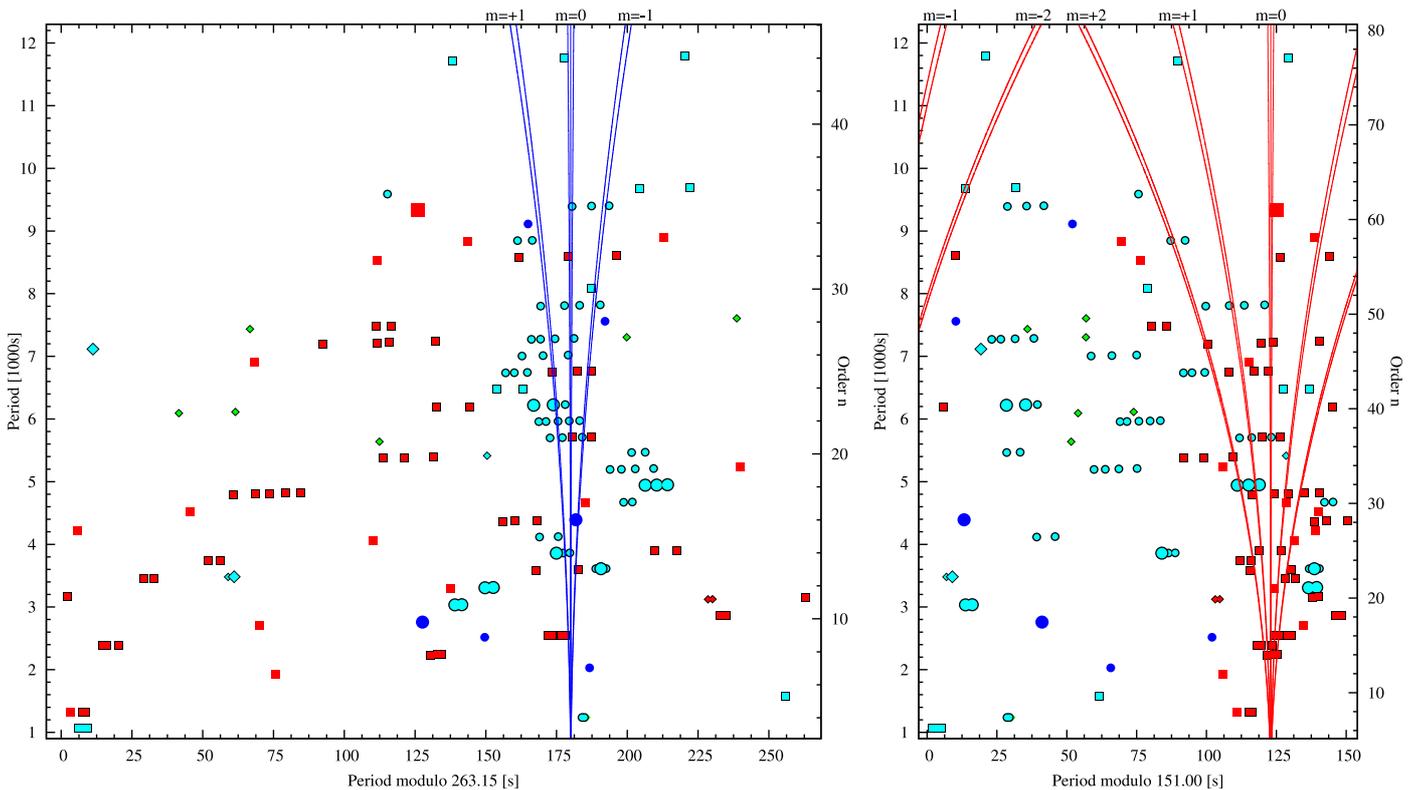}
\caption{\label{fig:echelle}
\'Echelle diagram for \ellone\ (left) and \elltwo\ (right).
Detected modes are marked according to their period on the abscissa, with a
cyclic folding on the asymptotic period on the ordinate axis.
The right-hand axis gives the order $n$ of the mode, according to the
asymptotic relation (Eq.\,\ref{eq:asym}).
Blue circles mark modes identified as \ellone, with outlined cyan circles
indicating those that appear in multiplets. Red squares mark \elltwo\ modes,
again with outlined points indicating multiplets.
Cyan diamonds mark trapped \ellone\ modes, red diamonds mark trapped \elltwo\ modes,
and green diamonds mark modes that do not fit either sequence. Outlined cyan
squares indicate modes that can be either \ellone\ or \elltwo. The high-amplitude
peaks ($A$\,$>$\,250\,ppm) in the full dataset are marked with enlarged symbols.
}
\end{figure*}

\subsection{The asymptotic period spacing}

In the asymptotic limit of stellar pulsation theory, consecutive
$g$ modes follow the relation
\begin{equation}\label{eq:asym}
P_{\ell,n} = \frac{\mit\Pi_0}{L} n + \epsilon_\ell
\end{equation}
where $L$\,=\,$\sqrt{\ell(\ell+1)}$, $\mit\Pi_0$ is the reduced period
spacing in the asymptotic limit, and $\epsilon_\ell$ is a constant
offset for each $\ell$.

A hallmark of the V1093\,Her stars revealed by \kep\ is that the
asymptotic period relation is readily detectable by the period
spacings, \DP\ of the modes \citep{reed11c,reed12a}. The favoured method for
determining the mean spacing is the Kolmogorov-Smirnov (KS) test,
which produces high negative values at the most frequently observed
spacing in a dataset.
Figure\,\ref{fig:kstest} shows the KS statistic for two period
lists, the full set listed in Table\,\ref{tbl:freqs} and a
list truncated to contain only modes with amplitudes higher than
1000\,ppm. A clear minimum is seen around
260\,s in both sets.  Only for the list including the low-amplitude modes
does the KS test show a second
minimum at \DP\,=\,150\,s. The 1/$\sqrt{3}$ relationship between these
two peaks is the signature of the period difference between 
\ellone\ and \elltwo\ modes in the asymptotic approximation.

By plotting the $P$ modulo \DP\ versus $P$ for the two period
spacings, one can construct an \'echelle diagram for $g$-mode pulsators.
Unlike the \'echelle diagram used for $p$-mode pulsators,
which are evenly spaced in frequency and have the same spacing, $\Delta\nu$,
for different orders $\ell$, the $g$-mode \'echelle diagram must be folded
on a different \DP\ for each $\ell$. Figure\,\ref{fig:echelle} shows the
\'echelle diagrams for \ellone\ and \elltwo\ for the \npk\ detected peaks
in \hamfast.
After starting with the \DP\ detected by the KS test, we made some iterations
of identifying peaks and adjusting the spacing slightly until a reasonable
picture emerged. 

In the \ellone\ \'echelle diagram, one can clearly see a
ridge of modes that meanders around the vertical lines drawn at
$\epsilon$\,=\,180\,s. The vertical lines curve to indicate
how the frequency splitting for \emone\ translate into period space.
Two lines are drawn for each $m$ to indicate the frequency resolution of
the full dataset. The \ellone\ ridge includes almost all the most powerful
modes, but a few are significantly off either sequence.

\subsection{Assigning modes to observed periods}\label{sect:modeID}

While some mode identifications are obvious based on multiplet structures,
the deviations from a clear asymptotic sequence imply substantial
ambiguities in the labelling. The identifications listed in Table~\ref{tbl:freqs}
should therefore not be taken as anything more than what the authors consider to
be the most likely ones.
Also, some frequencies listed are clearly spurious, caused by amplitude variability,
since in several cases groups of four, or even five, peaks are listed as
belonging to one \ellone\ triplet. In some cases the sequence for \ellone\ overlaps
with the one of \elltwo, causing further ambiguity. For instance, in the case
of $f_{152-155}$ the two complex peaks (see Fig.\,\ref{fig:mainpk2}) are
likely to be $\ell,n$\,=\,1,35 and 2,61, with the \elltwo\ mode at
$\sim$107\,\uHz. Several other complex groups in Fig.\,\ref{fig:mainpk2}
are identified as superpositions of modes of different orders.
In some cases, a particular group might be either \ellone\ or \elltwo, but
has been identified based on the frequency splitting or because the 
identification for a given $\ell,n$ has already been assigned to a suitable
mode.

While the majority of frequencies in Table~\ref{tbl:freqs} can be identified
in this scheme, some clearly fall well off either sequence. If these
were all low-amplitude modes, we could dismiss them as $\ell$\,=\,3 or
higher, but two of them appear among the highest amplitude peaks plotted
in Fig.\,\ref{fig:mainpk}.
The third strongest peak is the single and stable $f_{123}$ at 140.5\,\uHz,
which sits way out on the left edge in Fig.\,\ref{fig:echelle}.
It falls between the relatively low-amplitude multiplets assigned ID's
$\ell,n$\,=\,1,26 and 1,27, which both match the \ellone\ sequence well,
and the nearest \elltwo\ mode, $n$\,=\,47, is also occupied.
A similar problem appears with the pair $f_{48-49}$, which is ranked as the
8$^{\mathrm th}$ highest mode in amplitude. The pair appears with a splitting
of 0.171\,\uHz, which is just slightly wider than that of the main
\mainmode\ triplet.
But it falls between two other higher-amplitude triplets assigned
$\ell,n$\,=\,1,12 and 1,13. The spacing between those triplets is 302\,s,
so higher than the average period spacing, but much too tight to permit another
\ellone\ mode to squeeze in. It is also sandwiched between two low-amplitude
pairs assigned $\ell,n$\,=\,2,22 and 2,23.
The remaining peaks that defy assignment in the asymptotic interpretation
are all low amplitude and might well be $\ell$\,$>$\,2.

\subsection{The trapped modes}

The most plausible interpretation of the off-sequence,
high-amplitude peaks is that they
are trapped modes, which are produced mostly by the H/He transition in the stratified
envelope as predicted by classical sdB models \citep{charpinet00}.
To visualise the trapping signature, theoretical papers often show a
period-spacing diagram where the period difference between consecutive modes are plotted
against period. When reduced period, $\mit\Pi$\,=\,$PL$, is used, modes with the
same radial order, $k$, should overlap in this diagram. But to compute the
required period differences, 
\DPr\,=\,$\mit\Pi_{k}$\,--\,$\mit\Pi_{k-1}$, we must have completely uninterrupted
sequences.
Note that when mode trapping occurs, extra modes are inserted into the asymptotic
sequence, so that the real number of radial nodes in the star, $k$, is higher than
the asymptotic order $n$.

\begin{figure}[t!]
\includegraphics[width=\hsize]{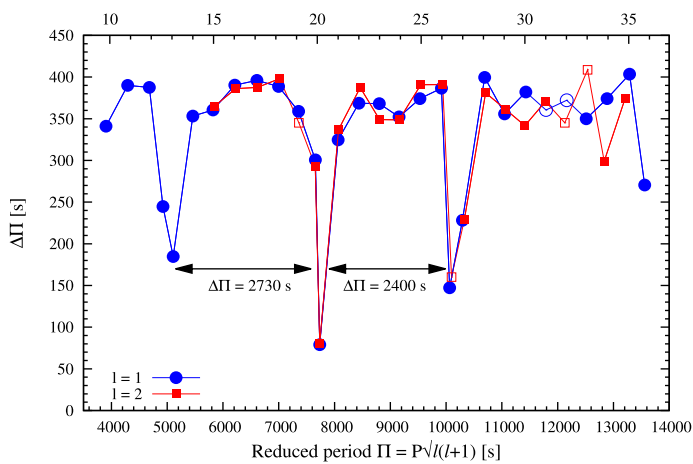}
\caption{
Period difference between consecutive modes of the \ellone\ and \elltwo\
sequences, after converting to reduced periods. The asymptotic order of
the modes, $n$, is indicated on the upper axis.
}
\label{fig:spacing}
\end{figure}

Inspecting the sequences of \ellone\ and \elltwo\ modes listed in Table~\ref{tbl:freqs},
including the trapped modes that are marked as $n$\,=\,t in the table,
we see that just a few modes are missing. To make a period-spacing diagram with
an uninterrupted sequence of consecutive modes, we must check each case and see if
we can find a suitable number to use in the sequence. Note that
due to the huge number of independent frequencies in the full FT of
the whole \kep\ dataset, we have maintained a 5-$\sigma$ significance threshold
in order to avoid too many spurious frequencies. However, when looking at a specific
frequency region suspected of containing a real frequency, it is justified to consider
a 4-$\sigma$ signal to be a significant detection.

The six modes needed to complete the sequences are
\ellone, $n$\,=\,31, 32, \elltwo, $n$\,=\,19, 32, 33,
and an \elltwo\ mode corresponding to the trapped \ellone\ mode
between $n$\,=\,26 and 27.
For \ellone, $n$\,=\,31 there is a feature at the 4-$\sigma$ level that we can use 
to complete the sequence of modes needed to construct the reduced period
diagram.
The same is the case for \elltwo, $n$\,=\,19 and 33.
The \ellone, $n$\,=\,32 mode should be at the position where the oddly spaced
\elltwo\ multiplet $f_{143-146}$ is found, so we use the highest of these peaks to
complete the sequence. A similar case can be made for \elltwo, $n$\,=\,32, which
should occur just where the highest peak in the FT is found, at $\sim$202\,\uHz.
And an \elltwo\ mode to correspond with the trapped \ellone\ mode at $\mit\Pi$\,=\,10064\,s
would be located at 4109\,s, which is in the region where
another strong \ellone\ mode, $f_{63}$, is seen.

After discovering the presence of trapped modes in \hamfast, we revisited the
table of frequencies to see if we could find evidence of other less obvious trapped
modes. The only feature we could find that seems reasonable to interpret
as a trapped mode is located between $n$\,=\,19 and 20. The two modes
$f_{38-39}$ consist of a relatively high-amplitude stable peak with a noisy
companion (shown in the first panel of Fig.~\ref{fig:mainpk2}).
It was first interpreted as \ellone, but there are no missing \ellone\
modes in this region. As a trapped mode, it could be either \ellone\ or
\elltwo. An \ellone\ interpetation would require a corresponding \elltwo\ mode
at $\sim$1804\,s, but this region is clean. An \elltwo\ interpretation 
requires the corresponding \ellone\ mode to
have an observed period of $\sim$5412\,s, at which we already find $f_{90}$.
This mode was first interpreted
as the fourth component of an incomplete \elltwo\ mode,
even though the splitting did not match the expected one very well.
Identifying $f_{90}$ as the trapped \ellone\ mode corresponding to a trapped
\elltwo\ mode provides a closer match than it did as part of an \elltwo\ multiplet.
We therefore adopt this latter interpretation, but note that the identification
for this trapped mode is not as clear as for the other two.

After completing the sequences by filling in for the six missing modes
and the additional trapping feature between $n$\,=\,19 and 20,
the reduced-period diagram shown in Fig.\,\ref{fig:spacing} emerges.
The modes identified in Table\,\ref{tbl:freqs} are marked with filled symbols
and the missing modes with open symbols in the figure.
All the observed multiplets have been reduced to a single period
for this figure. For the triplets and the three clear
\elltwo\ multiplets we have inferred the position of the central 
component. For other modes, the case is much more ambiguous. In general,
unless the period spacing could be interpreted in such a way as to locate the
correct centre, we simply used the highest peak. At low $n$ the error
this produces is tiny, but at high $n$ the error can be quite large.
The curves plotted in the \'echelle diagrams and labelled on the top axis
with the respective $m$, reflects this effect. 
The offset between the \ellone\ and 2 sequences in Fig.\,\ref{fig:spacing}
increases at high $n$, which is most likely caused by such mode
ambiguity rather than any physical effects.

The expected correspondence between the \ellone\ and \elltwo\ sequences
is striking. The diagram reveals three clear trapping features, where
at least one is clearly present in both the \ellone\ and \elltwo\ sequences.
The first is at low radial order where the \elltwo\ sequence is not present,
and the last includes a `missing' \elltwo\ point.
The difference in reduced period between consecutive trapped modes can be
seen to be $\mit\Pi_H$\,=\,$\sim$2400 and $\sim$2730\,s as
indicated by the horizontal arrows in Fig.\,\ref{fig:spacing}.

This observational reduced period diagram 
bears a striking resemblance to similar diagrams produced
from theoretical models, e.g.~Fig.\,3 in \citet{charpinet02a}.
The properties of this trapping structure depends
on the mass of the H-rich envelope and the position of the
transition zones inside the star \citep{charpinet00}.
Both the transition zone between the hydrogen and helium layers in the envelope
and the transition between the helium layer and the convective core, which
with time develops an increasing carbon-oxygen content and thus a higher
density, produce mode-trapping features that will affect the periods of
the trapped modes \citep{charpinet13a}.
Due to the complexities of these double trapping zones it
may not be straightforward to translate trapping period differences
into asteroseismic ages on the EHB.

\section{Conclusions}

We have analysed the complete \kep\ short-cadence lightcurve for \target,
and collected and analysed spectroscopic observations that reveal it to be
a system consisting of a V1093-Her pulsator orbiting a white dwarf.
When starting the frequency analysis of the pulsator, it soon became evident
that a large number of the observed periodicities fitted
neatly on the asymptotic sequences for \ellone\ and \elltwo, which is
similar to what we have seen with the other V1093-Her pulsators in the
\kep\ field. However, when we realised that a few of the main modes were clearly
incompatible with the asymptotic relation, we were immediately intrigued.
By accepting two high-amplitude modes as trapped \ellone\ modes, we
were finally able to make a convincing case that mode trapping, as predicted
by all theoretical models for V1093-Her pulsators, is present and detectable in
the \kep\ observations. Thanks to almost perfectly complete sequences
of consecutive \ellone\ and \elltwo\ modes, we were able to, for the first time,
generate an observed period-spacing diagram that shows convincing evidence for mode
trapping.
It is somewhat surprising that such features have not been spotted in other
V1093-Her pulsators observed with \kep. But it was only the high amplitude
of two of the trapped modes, and the fact that all available \ellone\ modes
in the sequence were already assigned, that tipped us off to this feature.
For other pulsators, trapped modes may hide in the unassigned low-amplitude
modes. It might be worthwhile revisiting the full sample of V1093-Her stars
in light of this revelation.

Many \ellone\ and \elltwo\ modes appear as rotationally split multiplets
indicating rotational periods that range from 34.5 to 44.2\,days, with the
most convincing \ellone\ modes averaging to $\sim$43\,d and the best
\elltwo\ modes $\sim$39\,d. An accurate determination of the rotation rate
from the observed multiplet splittings would require knowledge of the
$C_{n\ell}$ values from asteroseismic models. Until this becomes available
we must be satisfied with the rough estimate
$P_{\mathrm{rot}}$\,=\,41$\pm$3\,d.
We also found that a series of clear \elltwo\ multiplets all had
the middle $m$\,=\,0 component suppressed, implying a pulsation axis
observed at close to 55\degr, which is the same value as would be
required for the mass function of the binary to be compatible with
a canonical sdB mass and a normal 0.6\,\msol white-dwarf.
Since \bell\ must have been the original primary
of the progenitor system, the orbit must have been rather wide in order
to allow it to complete its red-giant-branch evolution, followed by
an asymptotic-giant-branch stage that brought it into contact with
the main-sequence progenitor of the current primary, \hamfast.
The parameters of the \target\ system and others like it can be used
to constrain the possible binary-interaction scenarios that allow
enough angular momentum to be lost from the system to
result in the observed configuration.

\begin{acknowledgements}

The authors gratefully acknowledge the \kep\ team and everybody who
has contributed to making this mission possible.
Funding for the \kepmi\ is provided by the NASA Science Mission Directorate.
We also thank Prof.~Uli Heber for kindly providing the model grids
used for the LTE atmospheric analysis.

The research leading to these results has received funding from the European
Research Council under the European Community's Seventh Framework Programme
(FP7/2007--2013)/ERC grant agreement N$^{\underline{\mathrm o}}$\,227224
({\sc prosperity}), and from the Research Council of KU~Leuven grant agreement
GOA/2008/04.
Funding for this research was also provided by the US National Science Foundation
grants \#1009436 and \#1312869, and the Polish National Science Centre under
project N$^{\underline{\mathrm o}}$\,UMO-2011/03/D/ST9/01914.

The spectroscopic observations used in this work were
collected at the Nordic Optical Telescope ({\scriptsize NOT})
at the Observatorio del Roque de los Muchachos (ORM)
on La Palma, operated jointly by Denmark, Finland, Iceland,
Norway, and Sweden; the William Herschel Telescope (WHT)
also at the ORM and operated by the Isaac Newton Group;
and the Mayall Telescope of Kitt Peak National Observatory,
which is operated by the Association of Universities for Research
in Astronomy under cooperative agreement with the NSF.

\end{acknowledgements}
\bibliographystyle{aa}
\bibliography{sdbrefs}

\onecolumn
\begin{appendix}
\section{Radial velocity data}
\begin{table*}[h]
\small\rm
\caption[]{Radial velocity data.}
\label{tbl:rvs}\centering
\begin{tabular}{crrl|crrl}
\hline\hline \noalign{\smallskip}
 HJD--2450000 & RV$_{\mathrm{obs}}$ & $\sigma_{\mathrm{RV}}$ & Obs & 
 HJD--2450000 & RV$_{\mathrm{obs}}$ & $\sigma_{\mathrm{RV}}$ & Obs \\ 
\noalign{\smallskip} \hline \noalign{\smallskip}
5434.44457 & 100.48 & 5.17 & WHT & 
6035.68704 & 1.28 & 14.40 & NOT \\ 
5434.56403 & 76.34 & 8.77 & WHT & 
6043.70719 & 103.06 & 12.55 & NOT \\ 
5435.51427 & -15.01 & 5.62 & WHT & 
6063.58724 & 112.58 & 17.04 & NOT \\ 
5436.43638 & 47.40 & 6.17 & WHT & 
6077.61366 & 104.05 & 14.75 & NOT \\ 
5437.46242 & 118.17 & 6.37 & WHT & 
6078.62528 & 18.37 & 14.49 & NOT \\ 
5438.44936 & 17.78 & 7.24 & WHT & 
6081.59025 & 64.40 & 16.16 & NOT \\ 
5439.58395 & 24.16 & 9.00 & WHT & 
6144.57627 & 70.06 & 16.18 & NOT \\ 
5712.58996 & 30.79 & 13.70 & NOT & 
6205.52762 & 27.09 & 24.05 & NOT\tablefootmark{a} \\ 
5712.59864 & 54.04 & 20.66 & NOT & 
6206.54680 & 80.30 & 21.87 & NOT\tablefootmark{a} \\ 
5712.70991 & 28.00 & 13.08 & NOT & 
6207.51302 & -15.79 & 19.45 & NOT\tablefootmark{a} \\ 
5720.64129 & 10.68 & 13.72 & NOT & 
6208.55354 & -27.43 & 22.62 & NOT\tablefootmark{a} \\ 
5720.65791 & 22.18 & 12.95 & NOT & 
6201.64394 & 0.86 & 3.99 & KPNO \\ 
5765.51459 & 78.55 & 20.82 & NOT & 
6201.65153 & 4.85 & 4.71 & KPNO \\ 
5765.74243 & 87.53 & 18.47 & NOT & 
6201.81768 & 25.08 & 10.84 & KPNO \\ 
5766.38565 & 84.22 & 17.63 & NOT & 
6201.82499 & 27.15 & 8.94 & KPNO \\ 
5766.74516 & 62.02 & 23.42 & NOT & 
6199.61149 & 105.18 & 10.65 & KPNO \\ 
5999.67482 & 95.75 & 18.52 & NOT & 
6199.61904 & 110.11 & 7.89 & KPNO \\ 
5999.77523 & 109.96 & 14.12 & NOT & 
6200.61981 & 17.19 & 7.14 & KPNO \\ 
6026.67181 & 97.72 & 11.05 & NOT & 
6200.62717 & 11.46 & 7.91 & KPNO \\ 
6026.74829 & 98.88 & 13.02 & NOT & 
6200.75724 & 16.06 & 6.87 & KPNO \\ 
6027.75346 & 34.03 & 12.11 & NOT & 
6200.76454 & 22.55 & 8.33 & KPNO \\ 
6035.68704 & 1.28 & 14.40 & NOT & 
 & & & \\ 
\noalign{\smallskip} \hline
\end{tabular}
\tablefoot{
\tablefoottext{a}{Vertical slit}
}
\end{table*}

\section{Frequency data and figures}

\begin{longtab}
\centering\small\rm
\begin{longtable}{lrrrcrrl}
\caption{\label{tbl:freqs}Frequencies, periods, and amplitudes for \target.}\\
\hline\hline
   \multicolumn{1}{c}{ID} &
   \multicolumn{1}{c}{Frequency} &
   \multicolumn{1}{c}{Period} &
   \multicolumn{1}{c}{Amplitude} &
   \multicolumn{1}{c}{$\ell$} &
   \multicolumn{1}{c}{$n_1$} &
   \multicolumn{1}{c}{$n_2$} &
   \multicolumn{1}{l}{State} \\
   &
   \multicolumn{1}{c}{[\uHz]} &
   \multicolumn{1}{c}{[s]} &
   \multicolumn{1}{c}{[ppm]} & & & & \\
\hline
\endhead
\hline
\endfoot
   $f_{1}$ & 4071.478 &  245.61 &   31.1 & -- & -- & -- & noisy \\
   $f_{2}$ & 3073.775 &  325.33 &   65.2 & 1 & 1 & -- & appearing \\
   $f_{3}$ & 3073.179 &  325.40 &   62.0 & 1 & 1 & -- & rising \\
   $f_{4}$ & 3059.768 &  326.82 &   41.2 & 1 & 1 & -- & stable \\
   $f_{5}$ & 1612.946 &  619.98 &   27.1 & 1 & 2 & -- & noisy \\
   $f_{6}$ &  944.384 & 1058.89 &   46.4 & 1|2 & 3 & 6 & disappearing \\
   $f_{7}$ &  943.896 & 1059.44 &   44.2 & 1|2 & 3 & 6 & appearing \\
   $f_{8}$ &  942.921 & 1060.53 &   62.5 & 1|2 & 3 & 6 & appearing \\
   $f_{9}$ &  942.335 & 1061.19 &   51.4 & 1|2 & 3 & 6 & disappearing \\
  $f_{10}$ &  941.646 & 1061.97 &   47.2 & 1|2 & 3 & 6 & messy \\
  $f_{11}$ &  808.674 & 1236.59 &   74.9 & 1 & 4 & -- & drifting \\
  $f_{12}$ &  808.117 & 1237.44 &   47.9 & 1 & 4 & -- & drifting \\
  $f_{13}$ &  807.691 & 1238.10 &   41.8 & -- & -- & -- & disappearing \\
  $f_{14}$ &  758.180 & 1318.95 &   29.2 & 2 & -- & 8 & noisy \\
  $f_{15}$ &  755.756 & 1323.18 &  126.1 & 2 & -- & 8 & drifting \\
  $f_{16}$ &  755.449 & 1323.72 &   66.0 & 2 & -- & 8 & drifting \\
  $f_{17}$ &  755.132 & 1324.27 &   48.3 & 2 & -- & 8 & drifting \\
  $f_{18}$ &  636.289 & 1571.61 &   44.1 & 1|2 & 5 & 10 & stable \\
  $f_{19}$ &  521.456 & 1917.71 &   27.5 & 2 & -- & 12 & noisy \\
  $f_{20}$ &  492.927 & 2028.70 &   86.7 & 1 & 7 & -- & stable \\
  $f_{21}$ &  447.306 & 2235.60 &  108.7 & 2 & -- & 14 & stable \\
  $f_{22}$ &  446.789 & 2238.19 &  119.3 & 2 & -- & 14 & stable \\
  $f_{23}$ &  446.530 & 2239.49 &   42.5 & 2 & -- & 14 & stable \\
  $f_{24}$ &  419.636 & 2383.02 &   95.5 & 2 & -- & 15 & stable \\
  $f_{25}$ &  419.389 & 2384.42 &   67.2 & 2 & -- & 15 & dropping \\
  $f_{26}$ &  418.646 & 2388.65 &   71.5 & 2 & -- & 15 & stable \\
  $f_{27}$ &  397.141 & 2518.00 &   98.9 & 1 & 9 & -- & stable \\
  $f_{28}$ &  393.633 & 2540.44 &   82.5 & 2 & -- & 16 & stable \\
  $f_{29}$ &  393.398 & 2541.96 &  115.7 & 2 & -- & 16 & stable \\
  $f_{30}$ &  392.931 & 2544.98 &  153.6 & 2 & -- & 16 & stable \\
  $f_{31}$ &  392.696 & 2546.50 &   53.3 & 2 & -- & 16 & dropping \\
  $f_{32}$ &  370.140 & 2701.68 &   29.0 & 2 & -- & 17 & noisy \\
  $f_{33}$ &  362.434 & 2759.12 &  518.1 & 1 & 10 & -- & stable \\
  $f_{34}$ &  349.136 & 2864.21 &   29.5 & 2 & -- & 18 & noisy \\
  $f_{35}$ &  348.878 & 2866.33 &   59.0 & 2 & -- & 18 & appearing \\
  $f_{36}$ &  329.629 & 3033.71 &  260.1 & 1 & 11 & -- & stable \\
  $f_{37}$ &  329.365 & 3036.15 &  384.5 & 1 & 11 & -- & stable \\
  $f_{38}$ &  320.182 & 3123.22 &   44.7 & 2 & -- & t & appearing \\
  $f_{39}$ &  320.025 & 3124.75 &  207.9 & 2 & -- & t & stable \\
  $f_{40}$ &  316.678 & 3157.78 &   50.7 & 2 & -- & 20 & appearing \\
  $f_{41}$ &  316.450 & 3160.05 &   29.0 & 2 & -- & 20 & noisy \\
  $f_{42}$ &  303.458 & 3295.35 &   68.0 & 2 & -- & 21 & disappearing \\
  $f_{43}$ &  302.341 & 3307.53 &  480.8 & 1 & 12 & -- & stable \\
  $f_{44}$ &  302.209 & 3308.96 &  150.0 & 1 & 12 & -- & dropping \\
  $f_{45}$ &  302.076 & 3310.43 &  403.5 & 1 & 12 & -- & stable \\
  $f_{46}$ &  289.843 & 3450.15 &   44.7 & 2 & -- & 22 & appearing \\
  $f_{47}$ &  289.538 & 3453.78 &   89.0 & 2 & -- & 22 & appearing \\
  $f_{48}$ &  287.361 & 3479.94 &   67.3 & 1 & t & -- & stable \\
  $f_{49}$ &  287.190 & 3482.01 &  279.8 & 1 & t & -- & stable \\
  $f_{50}$ &  278.665 & 3588.54 &   47.7 & 2 & -- & 23 & disappearing \\
  $f_{51}$ &  277.512 & 3603.45 &   55.0 & 2 & -- & 23 & disappearing \\
  $f_{52}$ &  277.026 & 3609.77 &  180.8 & 1 & 13 & -- & rising \\
  $f_{53}$ &  276.885 & 3611.60 &  433.8 & 1 & 13 & -- & stable \\
  $f_{54}$ &  276.744 & 3613.45 &  130.4 & 1 & 13 & -- & stable \\
  $f_{55}$ &  267.629 & 3736.51 &   48.3 & 2 & -- & 24 & dropping, drifting \\
  $f_{56}$ &  267.373 & 3740.09 &   54.3 & 2 & -- & 24 & drifting \\
  $f_{57}$ &  259.132 & 3859.04 &  612.4 & 1 & 14 & -- & stable \\
  $f_{58}$ &  258.974 & 3861.40 &  207.7 & 1 & 14 & -- & stable \\
  $f_{59}$ &  258.814 & 3863.78 &   66.1 & 1 & 14 & -- & stable \\
  $f_{60}$ &  256.809 & 3893.94 &   47.9 & 2 & -- & 25 & disappearing \\
  $f_{61}$ &  256.295 & 3901.76 &   26.9 & 2 & -- & 25 & noisy \\
  $f_{62}$ &  246.460 & 4057.45 &   30.7 & 2 & -- & 26 & noisy \\
  $f_{63}$ &  242.939 & 4116.26 &  131.4 & 1 & 15 & -- & rising \\
  $f_{64}$ &  242.553 & 4122.81 &  224.9 & 1 & 15 & -- & stable \\
  $f_{65}$ &  237.174 & 4216.30 &   51.1 & 2 & -- & 27 & appearing \\
  $f_{66}$ &  229.010 & 4366.62 &   58.5 & 2 & -- & 28 & disappearing \\
  $f_{67}$ &  228.791 & 4370.81 &   51.3 & 2 & -- & 28 & disappearing \\
  $f_{68}$ &  228.378 & 4378.71 &   59.7 & 2 & -- & 28 & appearing \\
  $f_{69}$ &  227.675 & 4392.22 &  950.8 & 1 & 16 & -- & stable \\
  $f_{70}$ &  221.274 & 4519.28 &   46.3 & 2 & -- & 29 & appearing \\
  $f_{71}$ &  214.650 & 4658.74 &   32.6 & 2 & -- & 30 & noisy \\
  $f_{72}$ &  214.029 & 4672.27 &   68.3 & 1 & 17 & -- & rising \\
  $f_{73}$ &  213.891 & 4675.28 &   49.2 & 1 & 17 & -- & stable \\
  $f_{74}$ &  208.428 & 4797.81 &   56.2 & 2 & -- & 31 & dropping \\
  $f_{75}$ &  208.106 & 4805.24 &   74.2 & 2 & -- & 31 & dropping \\
  $f_{76}$ &  207.885 & 4810.36 &   77.8 & 2 & -- & 31 & disappearing \\
  $f_{77}$ &  207.647 & 4815.86 &   52.5 & 2 & -- & 31 & rising \\
  $f_{78}$ &  207.427 & 4820.98 &   67.4 & 2 & -- & 31 & rising \\
  $f_{79}$ &  202.306 & 4943.02 &  509.5 & 1 & 18 & -- & stable \\
  $f_{80}$ &  202.138 & 4947.10 &  577.2 & 1 & 18 & -- & stable \\
  $f_{81}$ &  201.983 & 4950.90 & 1367.5 & 1 & 18 & -- & stable \\
  $f_{82}$ &  192.532 & 5193.95 &   35.9 & 1 & 19 & -- & dropping \\
  $f_{83}$ &  192.390 & 5197.76 &   46.3 & 1 & 19 & -- & drifting \\
  $f_{84}$ &  192.215 & 5202.50 &   30.5 & 1 & 19 & -- & noisy \\
  $f_{85}$ &  191.962 & 5209.37 &   31.4 & 1 & 19 & -- & noisy \\
  $f_{86}$ &  190.849 & 5239.75 &   31.2 & 2 & -- & 34 & noisy \\
  $f_{87}$ &  185.994 & 5376.51 &   41.5 & 2 & -- & 35 & disappearing \\
  $f_{88}$ &  185.681 & 5385.58 &  110.3 & 2 & -- & 35 & rising \\
  $f_{89}$ &  185.377 & 5394.41 &   76.5 & 2 & -- & 35 & rising \\
  $f_{90}$ &  184.724 & 5413.48 &   64.2 & 1 & t & -- & appearing \\
  $f_{91}$ &  182.983 & 5464.97 &   40.7 & 1 & 20 & -- & disappearing \\
  $f_{92}$ &  182.838 & 5469.33 &   50.8 & 1 & 20 & -- & appearing \\
  $f_{93}$ &  177.357 & 5638.35 &   31.3 & -- & -- & -- & noisy \\
  $f_{94}$ &  175.473 & 5698.87 &  119.8 & 1 & 21 & -- & stable \\
  $f_{95}$ &  175.340 & 5703.21 &   89.2 & 1 & 21 & -- & stable \\
  $f_{96}$ &  175.243 & 5706.36 &  120.1 & 2 & -- & 37 & stable \\
  $f_{97}$ &  175.121 & 5710.35 &   41.6 & 1 & 21 & -- & rising \\
  $f_{98}$ &  175.024 & 5713.51 &   73.6 & 2 & -- & 37 & rising \\
  $f_{99}$ &  167.852 & 5957.62 &   96.4 & 1 & 22 & -- & rising \\
 $f_{100}$ &  167.770 & 5960.53 &   93.1 & 1 & 22 & -- & rising \\
 $f_{101}$ &  167.653 & 5964.69 &   67.3 & 1 & 22 & -- & dropping \\
 $f_{102}$ &  167.531 & 5969.05 &   92.7 & 1 & 22 & -- & rising \\
 $f_{103}$ &  167.440 & 5972.27 &  152.6 & 1 & 22 & -- & dropping \\
 $f_{104}$ &  164.094 & 6094.08 &   51.8 & -- & -- & -- & appearing \\
 $f_{105}$ &  163.562 & 6113.91 &   75.9 & -- & -- & -- & stable \\
 $f_{106}$ &  161.683 & 6184.94 &  248.3 & 2 & -- & 40 & rising, drifting \\
 $f_{107}$ &  161.391 & 6196.11 &   70.2 & 2 & -- & 40 & rising \\
 $f_{108}$ &  160.790 & 6219.27 &  404.2 & 1 & 23 & -- & rising \\
 $f_{109}$ &  160.625 & 6225.69 &  316.9 & 1 & 23 & -- & stable \\
 $f_{110}$ &  160.504 & 6230.39 &  110.7 & 1 & 23 & -- & rising \\
 $f_{111}$ &  154.573 & 6469.42 &   52.4 & 1|2 & 24 & 42 & appearing \\
 $f_{112}$ &  154.346 & 6478.95 &   62.7 & 1|2 & 24 & 42 & appearing \\
 $f_{113}$ &  148.462 & 6735.73 &   78.9 & 1 & 25 & -- & appearing \\
 $f_{114}$ &  148.395 & 6738.75 &  157.2 & 1 & 25 & -- & stable \\
 $f_{115}$ &  148.293 & 6743.41 &   53.0 & 1 & 25 & -- & dropping \\
 $f_{116}$ &  148.102 & 6752.09 &   54.9 & 2 & -- & 44 & appearing \\
 $f_{117}$ &  147.919 & 6760.47 &   34.0 & 2 & -- & 44 & stable \\
 $f_{118}$ &  147.789 & 6766.40 &   37.0 & 2 & -- & 44 & rising \\
 $f_{119}$ &  144.712 & 6910.28 &   29.7 & 2 & -- & 45 & noisy \\
 $f_{120}$ &  142.762 & 7004.67 &   73.3 & 1 & 26 & -- & appearing \\
 $f_{121}$ &  142.610 & 7012.15 &   84.2 & 1 & 26 & -- & rising \\
 $f_{122}$ &  142.430 & 7020.98 &   55.0 & 1 & 26 & -- & appearing \\
 $f_{123}$ &  140.522 & 7116.32 &  698.9 & 1 & t & -- & dropping \\
 $f_{124}$ &  138.937 & 7197.49 &  118.9 & 2 & -- & 47 & rising \\
 $f_{125}$ &  138.569 & 7216.62 &   72.1 & 2 & -- & 47 & disappearing \\
 $f_{126}$ &  138.493 & 7220.58 &   61.4 & 2 & -- & 47 & appearing \\
 $f_{127}$ &  138.169 & 7237.52 &   57.8 & 2 & -- & 47 & appearing \\
 $f_{128}$ &  137.529 & 7271.22 &   81.0 & 1 & 27 & -- & appearing \\
 $f_{129}$ &  137.469 & 7274.34 &  195.5 & 1 & 27 & -- & appearing \\
 $f_{130}$ &  137.385 & 7278.81 &   91.0 & 1 & 27 & -- & rising \\
 $f_{131}$ &  137.244 & 7286.29 &   86.0 & 1 & 27 & -- & stable \\
 $f_{132}$ &  136.895 & 7304.85 &   45.4 & -- & -- & -- & noisy \\
 $f_{133}$ &  134.501 & 7434.89 &   29.0 & -- & -- & -- & noisy \\
 $f_{134}$ &  133.700 & 7479.45 &   41.8 & 2 & -- & 49 & disappearing \\
 $f_{135}$ &  133.603 & 7484.84 &   44.4 & 2 & -- & 49 & appearing \\
 $f_{136}$ &  132.272 & 7560.15 &   63.9 & 1 & 28 & -- & appearing \\
 $f_{137}$ &  131.455 & 7607.14 &   47.4 & -- & -- & -- & stable \\
 $f_{138}$ &  128.194 & 7800.71 &   54.8 & 1 & 29 & -- & disappearing \\
 $f_{139}$ &  128.053 & 7809.28 &   48.8 & 1 & 29 & -- & appearing \\
 $f_{140}$ &  127.962 & 7814.84 &   48.0 & 1 & 29 & -- & appearing \\
 $f_{141}$ &  127.844 & 7822.05 &   42.8 & 1 & 29 & -- & disappearing \\
 $f_{142}$ &  123.733 & 8081.92 &   51.8 & 1|2 & 30 & 53 & disappearing \\
 $f_{143}$ &  117.197 & 8532.63 &   50.0 & 2 & -- & 56 & disappearing \\
 $f_{144}$ &  116.517 & 8582.45 &   64.2 & 2 & -- & 56 & appearing \\
 $f_{145}$ &  116.279 & 8600.01 &   70.3 & 2 & -- & 56 & rising \\
 $f_{146}$ &  116.048 & 8617.12 &   51.5 & 2 & -- & 56 & stable \\
 $f_{147}$ &  113.280 & 8827.72 &   46.5 & 2 & -- & 57 & appearing \\
 $f_{148}$ &  113.056 & 8845.20 &   34.8 & 1 & 33 & -- & noisy \\
 $f_{149}$ &  112.998 & 8849.74 &   40.0 & 1 & 33 & -- & disappearing \\
 $f_{150}$ &  112.400 & 8896.83 &   46.4 & 2 & -- & 58 & appearing \\
 $f_{151}$ &  109.745 & 9112.01 &   82.9 & 1 & 34 & -- & appearing \\
 $f_{152}$ &  107.120 & 9335.32 &  211.5 & 2 & -- & 61 & rising \\
 $f_{153}$ &  106.489 & 9390.67 &  135.6 & 1 & 35 & -- & appearing \\
 $f_{154}$ &  106.410 & 9397.62 &  102.2 & 1 & 35 & -- & stable, messy \\
 $f_{155}$ &  106.342 & 9403.64 &   49.2 & 1 & 35 & -- & appearing, messy \\
 $f_{156}$ &  104.290 & 9588.62 &   54.2 & 1 & 36 & -- & appearing \\
 $f_{157}$ &  103.334 & 9677.32 &   55.4 & 1|2 & 36 & 63 & rising, messy \\
 $f_{158}$ &  103.140 & 9695.56 &   40.2 & 1|2 & 36 & 63 & stable \\
 $f_{159}$ &   85.347 & 11716.84 &   48.3 & 1|2 & 44 & 76 & dropping \\
 $f_{160}$ &   85.060 & 11756.43 &   53.9 & 1|2 & 44 & 77 & stable \\
 $f_{161}$ &   84.755 & 11798.71 &   61.2 & 1|2 & 44 & 77 & drifting \\
 $f_{162}$ &   46.840 & 21349.31 &   43.6 & -- & -- & -- & appearing \\
\hline
\end{longtable}\end{longtab}

\begin{figure*}[t!]
\includegraphics[width=9cm]{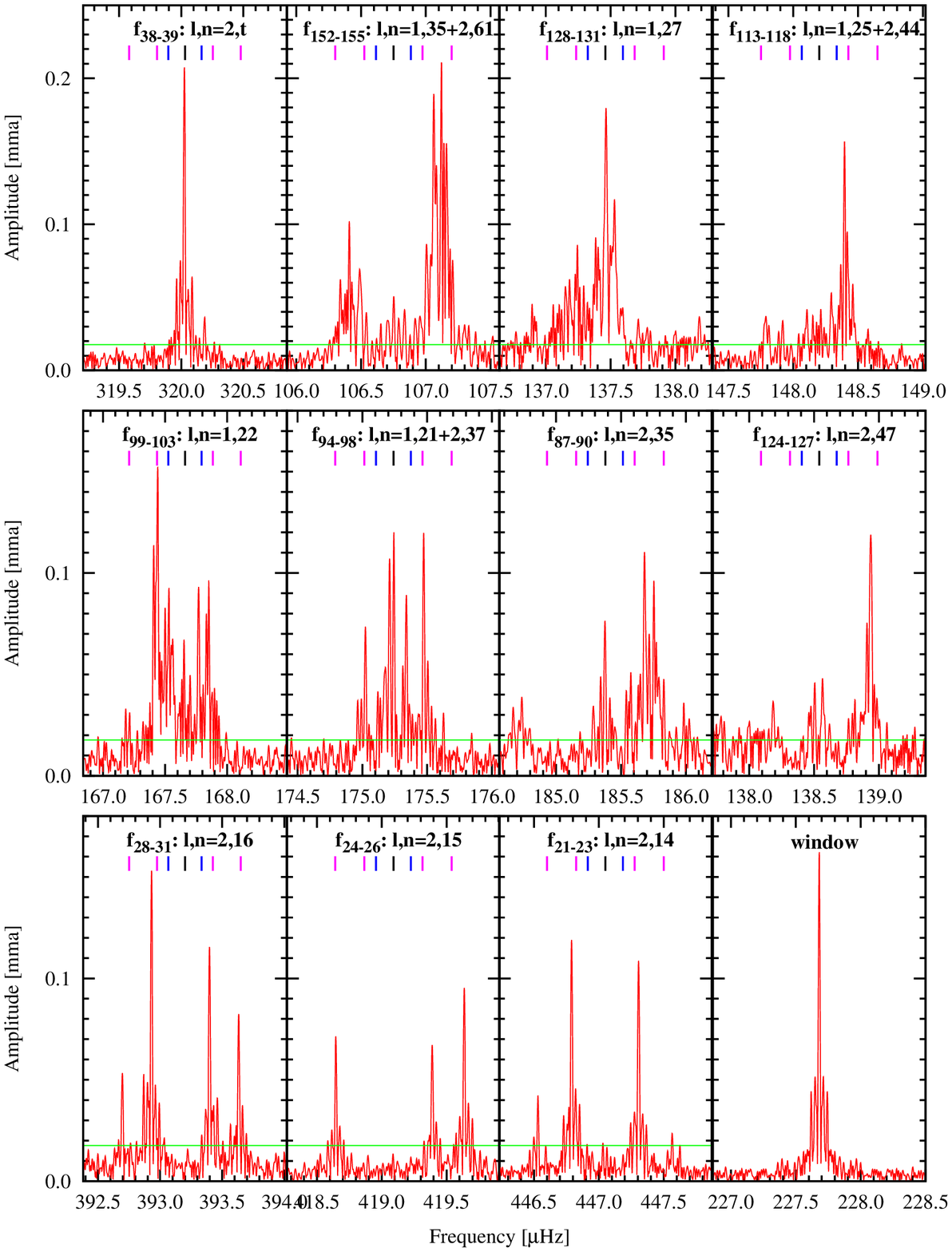}
\includegraphics[width=9cm]{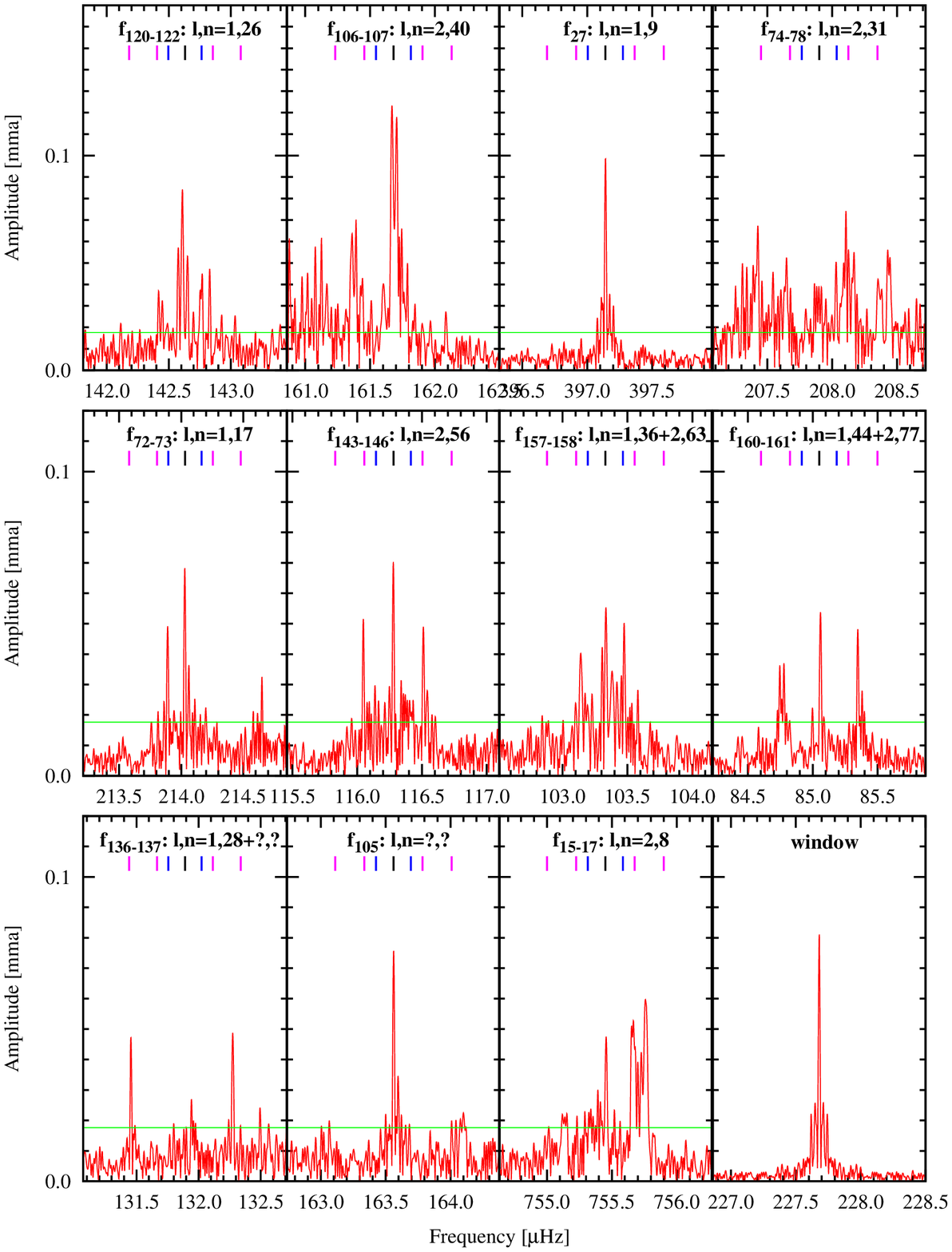}
\caption{
Some of the most significant peaks in the \hamfast\ Fourier spectrum, following
from the sequence shown in \ref{fig:mainpk}.
}
\label{fig:mainpk2}
\end{figure*}

\end{appendix}

\end{document}